\documentclass[12pt]{iopart}

\expandafter\let\csname equation*\endcsname\relax
\expandafter\let\csname endequation*\endcsname\relax
\usepackage{amsmath}
\usepackage{amsmath}
\usepackage{graphicx}
\usepackage{xcolor}
\begin{document}

\title[StatPhys29 contribution]
{Competition at the front of expanding populations}

\author{Sergio Eraso  \&  Mehran Kardar}

\address{Department of Physics, Massachusetts Institute of Technology, Cambridge, MA, USA}
\ead{kardar@mit.edu}
\vspace{10pt}
\begin{indented}
\item[]\today
\end{indented}

\begin{abstract}
{When competing species grow into new territory, the population is dominated by descendants of successful ancestors at the expansion front. Successful ancestry depends on both the reproductive advantage (fitness), as well as ability and opportunity to colonize new domains. We present a model that integrates both elements by coupling the classic description of one-dimensional competition (Fisher equation) to the minimal model of front shape (KPZ equation). Macroscopic manifestations of these equations are distinct growth morphologies controlled by expansion rates, competitive abilities, or spatial anisotropy. In some cases the ability to expand in space may overcome reproductive advantage in colonizing new territory. When new traits appear with accumulating mutations, we find that variations in fitness in range expansion may be described by the Tracy--Widom distribution.}
\end{abstract}

%\maketitle

\section{KPZ and related projects: a historical perspective (Kardar)}

The Kardar--Parisi--Zhang (KPZ) equation was introduced in 1986 with Yi-Cheng Zhang and Giorgio Parisi as a minimal continuum description of kinetic roughening of growing interfaces~\cite{KPZ1986}. In its simplest form, the KPZ equation for a height field $h(x,t)$ reads
\begin{equation}
\partial_t h(x,t) = \nu \nabla^2 h(x,t) + \frac{\lambda}{2} \left[\nabla h(x,t)\right]^2 + \eta(x,t),
\label{eq:kpz}
\end{equation}
where $\nu$ represents an effective smoothening (e.g. due to surface tension), $\lambda$ is a parameter characterizing the nonlinear growth (e.g. from projection along the local normal, see Fig.~\ref{fig:kpz_geometry}), and $\eta$ is a stochastic noise representing random deposition events. This deceptively simple equation has come to define a universality class that encompasses a wide variety of phenomena, ranging from fluid flow and crystal growth to directed polymers in random media and fluctuating hydrodynamics.
From its inception, the KPZ equation has served as a minimal model for describing how local, noisy, and asymmetric non-equilibrium processes give rise to emergent, large-scale behavior. Like the diffusion equation in equilibrium systems, KPZ has appeared across an astonishing range of contexts; from growing interfaces and traffic flow to quantum entanglement dynamics and evolutionary biology. Its power lies in its universality: the equation encapsulates generic features without reliance on microscopic detail.

Shortly after the introduction of KPZ, the connection to directed polymers in random media (DPRM) was made explicit by  the Cole--Hopf transformation to a diffusion equation with multiplicative noise~\cite{KPZ-DPRM1987}. In this picture, the height field $h(x,t)$ is related to the free energy of a polymer wandering in a random potential. This mapping opened the door to using tools from  statistical mechanics of disordered systems, and enabled my introduction of the replica method combined with Bethe Ansatz~\cite{ReplicaBethe1987}.

I would like to acknowledge the contributions of a remarkable group of collaborators that enabled further explorations of KPZ universality after I took up a faculty position at MIT. The first wave of such research involved Ernesto Medina and Terry Hwa, we further generalized renormalization group techniques to the KPZ equation with correlated noise~\cite{MedinaHwa1989}. Deniz Ertaş joined shortly thereafter, and we investigated coupled KPZ equations—an early precedent of competition on growing fronts~\cite{ErtasKardar1994}. Maya Paczuski contributed to studies of growth-induced roughening of crystalline facets~\cite{HwaPaczuski1991}. 
Yonathan Shapir initiated study of interference of quantum paths in disordered media~\cite{Shapir89},
culminating in extensions of KPZ scaling to strongly localized electrons~\cite{MedinaKardar92}. 
Collaborations with Larry Saul and Nick Read led to further generalizations to directed  waves in random media~\cite{SaulKardarRead1992}, while with Leon Balents, we explored a range of topics from delocalization transitions to the unusual $n \to 0$ limit in interacting fermion models~\cite{BalentsKardar1992a,BalentsKardar1992b,BalentsKardar1993}. 

Closer to the original motivation of kinetic roughening, the phase ordering and roughening of growing films were studied with Barbara Drossel in~\cite{DrosselKardar2000}, where coupling between order-parameter dynamics and surface growth led to rich pattern formation, reminiscent of results presented in the next section. We also worked on energy barriers in flux line systems~\cite{DrosselKardar1995a,DrosselKardar1995b}, winding angle statistics~\cite{DrosselKardar1996}. Thorsten Emig joined in studies involving Bethe ansatz techniques and mesoscopic disorder fingerprints~\cite{EmigKardar2000}. Together, these collaborations helped solidify KPZ’s place in the broader study of glassy, disordered, and driven systems.

Later collaborations with Rava da Silveira explored stochastic evolution equations with exactly solvable steady states~\cite{SilveiraKardar2003}, revealing new directions even within systems that appeared non-integrable. And it would be remiss not to mention Tim Halpin-Healy, who has been involved in extending and clarifying KPZ studies since its inception.

After several years focused on polymers, Casimir forces, and biological systems, I returned to KPZ-related research in the 2010s, this time motivated by both mathematical advances and biological applications. With Sherry Chu, we examined the effects of correlated noise on directed polymers~\cite{ChuKardar2016}, and later collaborated with David Nelson and Daniel Beller to study evolution in range expansions with competition at rough boundaries~\cite{ChuEtAl2019}.

Recent work has revisited KPZ from the lens of microbial competition and mutation accumulation. In a collaboration with Jordan Horowitz, we investigated how front roughness affects fixation and directed percolation~\cite{HorowitzKardar2019}. With Daniel Swartz, Hayoun Lee, and Kirill Korolev, we analyzed the interplay between morphology and competition in two-dimensional colony expansions~\cite{SwartzEtAl2023}. Most recently, with Lauren Li, we studied specialization dynamics at the front of expanding populations, where stochastic growth processes and evolutionary selection are deeply intertwined~\cite{LiKardar2023}.

Through  these developments, I remain indebted to the many students, postdocs, and collaborators who shaped this evolving body of work. Their creativity, insight, and energy have driven the KPZ equation far beyond its initial conception, transforming it into a universal structure at the heart of nonequilibrium statistical physics. The present contribution builds on this legacy by coupling KPZ-type front roughening to models of competition and evolution at expanding fronts, thereby contributing to the topic of biological range expansions and fitness landscapes.

\section{Growth and Competition at Expanding Fronts}

Understanding how populations  compete at growing fronts is essential for modeling biological range expansions: Individuals at the leading edge are primarily responsible for colonizing unoccupied territory, and their evolutionary fate is strongly shaped by both fitness and geometry. These fronts are inherently noisy and fluctuating, with competition dynamics often mediated by the structure and roughness of the interface.
Studies of range expansion were significantly influenced by experimental and conceptual work of Hallatschek \emph{et al.}~\cite{Hallatschek2007}, who demonstrated how competition at frontiers promotes strong genetic segregation. In these experiments, microbial populations expanding into unoccupied territory spontaneously formed mono-allelic sectors, even when competing variants had identical growth rates. This phenomenon established growing fronts as a natural laboratory for studying nonequilibrium fluctuations combined with evolutionary dynamics.

\subsection*{Neutral range expansion}
An interesting observation from these experiments is that random reproduction at the boundaries separating different \textit{neutral sectors} leads to anomalous, \textit{super-diffusive} fluctuations. These observations were subsequently reproduced and analyzed in numerical simulations of stepping-stone models~\cite{Korolev2010}, and in many related studies, establishing that the anomalous wandering of sector boundaries is intimately connected to the roughness of the expanding front.

\begin{figure}[t]
\centering
\includegraphics[width=0.75\textwidth]{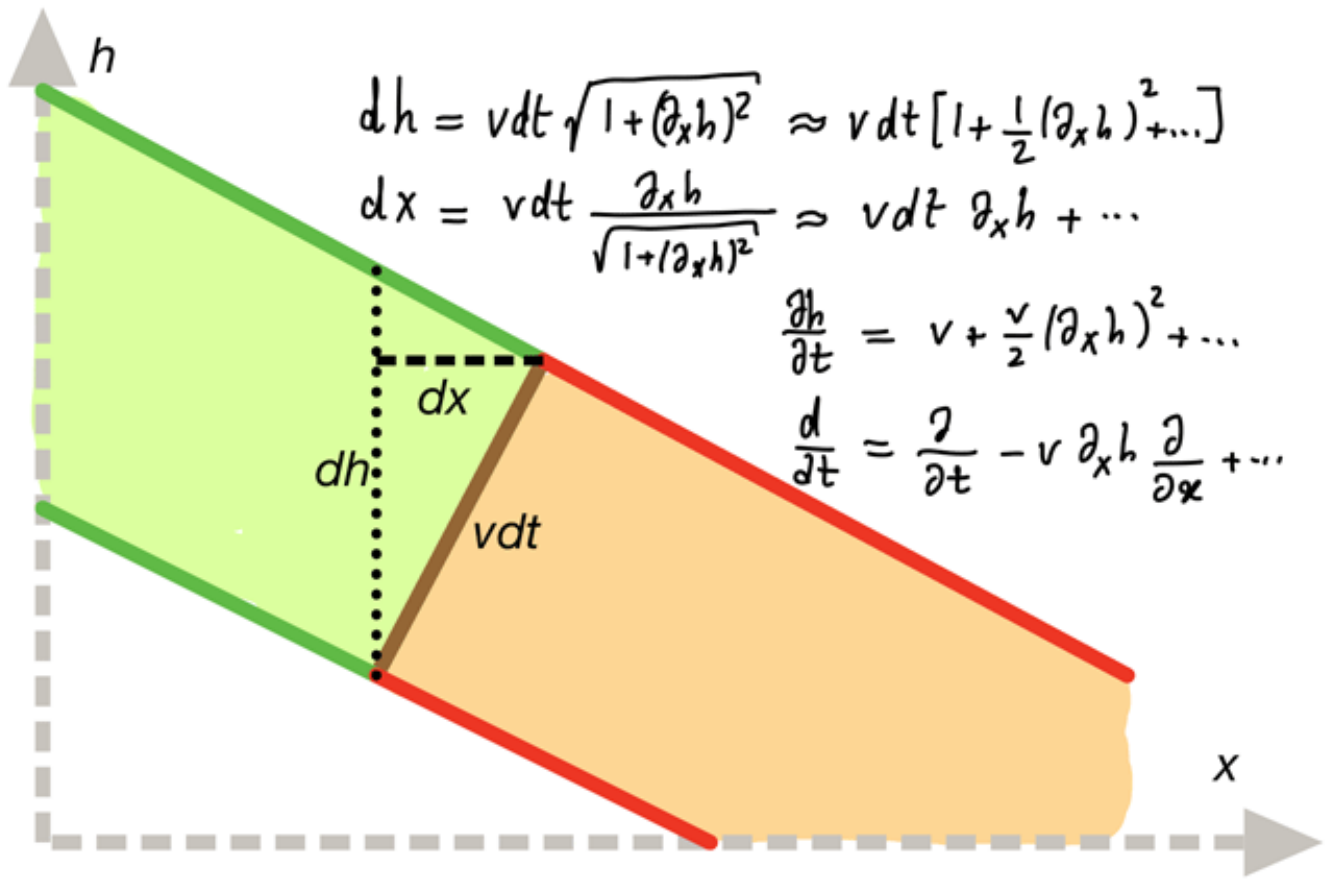}
\caption{Geometric origin of the nonlinear term in the KPZ equation when growth is always normal to the surface. There is a corresponding drift of a neutral sector boundary proportional to the local  slope.}
\label{fig:kpz_geometry}
\end{figure}

This connection naturally motivates a description of the growth front in terms of a fluctuating height field $h(x,t)$. The KPZ equation~\eqref{eq:kpz} provides the natural description of the evolving profile,  written here in the form
\begin{equation}
\frac{\partial h(x,t)}{\partial t}
=
v_0
+ \frac{v_0}{2}(\nabla h)^2
+ \nu \nabla^2 h
+ \eta(x,t),
\label{eq:kpz_form}
\end{equation}
where $v_0$ is the mean front velocity. As depicted in Fig.~\ref{fig:kpz_geometry}, in an isotropic system geometric considerations lead to the identification $\lambda=v_0$. The figure also indicates that isotropic growth leads to a drift of a neutral sector boundary located at position $x(t)$  according to
\begin{equation}
\dot{x}(t) = - v_0 \partial_x h(x,t).
\label{eq:boundary_drift}
\end{equation}
Thus, stochastic fluctuations in the height profile directly induce stochastic motion of the sector boundary.

Using Eq.~(\ref{eq:boundary_drift}), the mean-square displacement of the boundary satisfies
\begin{equation}
\langle x^2(t) \rangle
=
v_0^2
\int_0^t \!\! dt_1 \int_0^t \!\! dt_2
\,
\langle \partial_x h(t_1)\partial_x h(t_2) \rangle
\propto t^{2/z},
\end{equation}
where $z$ is the dynamic scaling exponent governing front fluctuations. The last identity follows from the scaling form
\begin{equation}
\langle [h(x,t)-h(0,0)]^2\rangle =
\begin{cases}
|x|\,\phi\!\left(x^2/t\right), & z=2 \quad \text{(Edwards--Wilkinson)},\\[6pt]
|x|\,\phi\!\left(x^{3/2}/t\right), & z=3/2 \quad \text{(KPZ in $d=1$)}.
\end{cases}
\label{eq:scaling}
\end{equation}
Consequently, in the KPZ universality class one obtains $\langle x^2(t)\rangle \sim t^{4/3}$, in agreement with experiments and simulations.

\begin{figure}[t]
\centering
\includegraphics[width=0.75\textwidth]{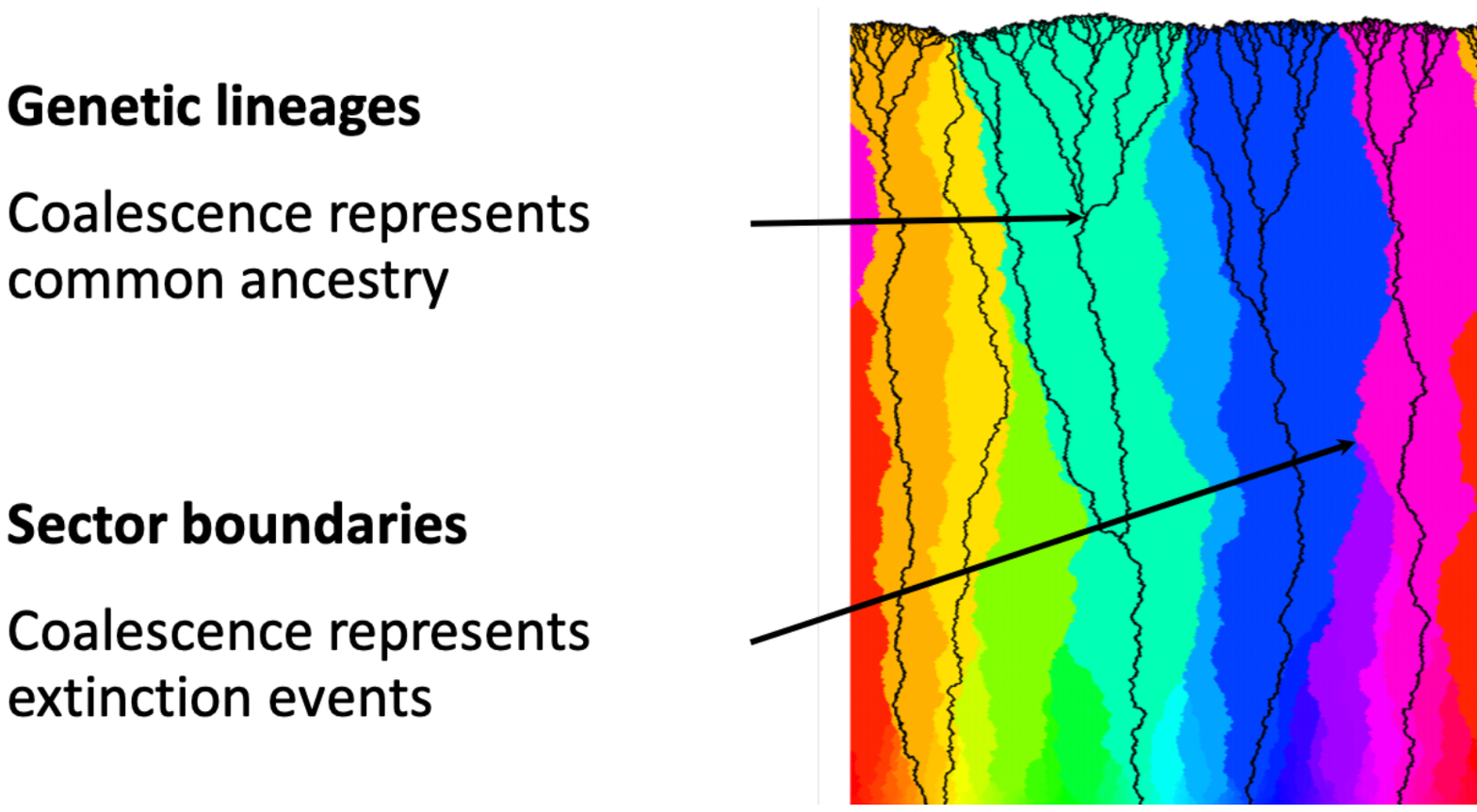}
\caption{Consequences of super-diffusive boundary motion: The colors indicate different (neutral)
species seeded at the bottom. Random reproduction events cause fluctuations of the sector boundaries. Coalescence of sector boundaries removes the enclosed species. The black lines trace back the ancestry of the individuals at the final front. Coalescence of these lines indicates a common ancestor.}
\label{fig:lca}
\end{figure}

An important evolutionary consequence of this super-diffusive wandering was noted in~\cite{ChuEtAl2019}: enhanced boundary fluctuations accelerate the loss of genetic diversity, leading to more recent and fewer last common ancestors, as depicted in Fig.~\ref{fig:lca}.

\subsection*{Competitive range expansion}
The experiments and analysis described above pertain to competition between \emph{neutral} species with identical growth rates. In general, however, competing strains differ in their reproductive success, and selection must be incorporated explicitly. If reproduction events can be ignored in the bulk of the film (due to space or nutrient constraints), it becomes possible to describe its evolution in terms of two coupled fields: the front height $h(x,t)$ and the local fraction $f(x,t)$ of one strain. Symmetry constraints then relate the evolution of these two fields to the KPZ equation, coupled to the Fisher--Kolmogorov--Petrovsky--Piscounov (Fisher--KPP) equation, a paradigmatic model of invasion fronts. The coupled equations take the form
\begin{align}
\partial_t h &= 
 v_0 + \frac{v_0}{2}(\nabla h)^2 +\nu \nabla^2 h
+ \alpha f,
\label{eq:coupled_h}\\
\partial_t f &=
s(f)\,(1-f)
+ D \nabla^2 f
+ v_0 \nabla f \cdot \nabla h,
\label{eq:coupled_f}
\end{align}
where $\alpha$  represents  the modification of the local front velocity by $\alpha$ as the fraction of the mutant species is increased from 0 to 1. The final term describes the drift of the sector boundary with slope  $\nabla h$, consistent with Eq.~(\ref{eq:boundary_drift}) and Fig.~\ref{fig:kpz_geometry}.

We first consider the simple case $s(f)=s_0 f$, with $s_0$ indicating the reproductive (fitness) advantage of the mutant species. Even in this limit, the coupled equations can lead to distinct morphologies as the parameters $\alpha$  and  $s_0$ are varied.
\begin{figure}[t]
\centering
\includegraphics[width=0.75\textwidth]{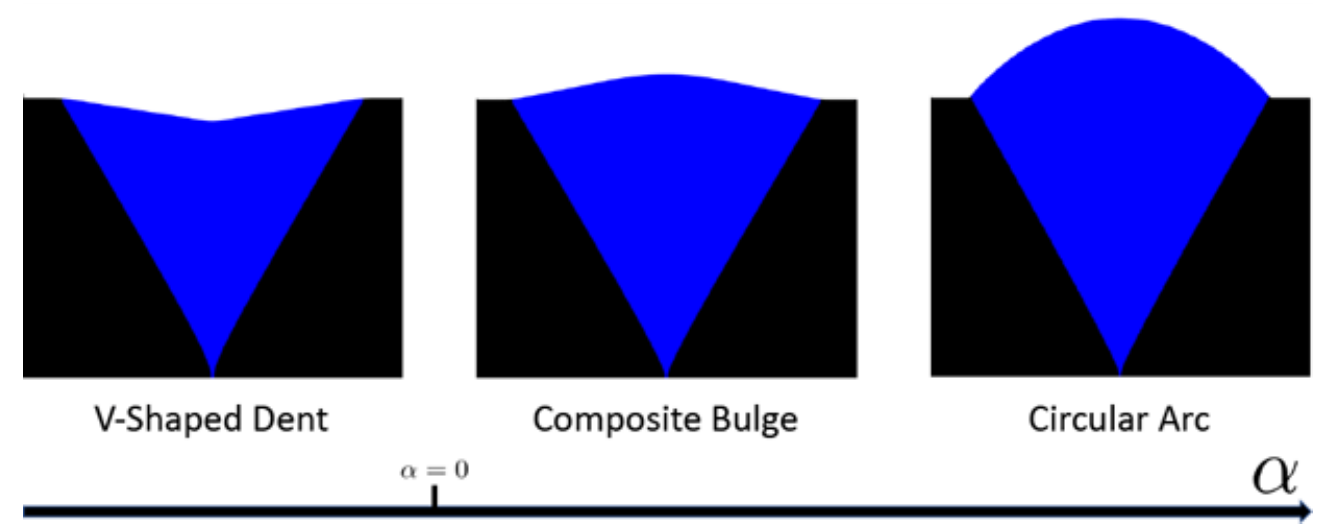}
\caption{Possible morphologies of competing fronts for $s(f)=s_0f$: circular arc, composite bulge, and V-shaped dent.}
\label{fig:morphologies}
\end{figure}
\begin{itemize}
\item Large values of $\alpha$ cause the mutant sector to bloom out in a \textit{circular-arc} morphology. Flat profiles $h(x,t|f=0)=v_0t$, now bound the circular arc with 
\begin{equation}\label{eq:parabola}
    h(x,t|f=1) = (v_0+\alpha)t - \frac{x^2}{2v_0 t}\,.
\end{equation}
Note that the parabolic form is a solution to $\partial_t h = v_0 +{v_0}(\nabla h)^2/2 + \alpha $,
representing the top of a circular arc. Geometric considerations indicate that the circular arc region expands into the flat portions with invasion speed
\begin{equation}
u = \sqrt{2\alpha v_0}
\approx \sqrt{(v_0+\alpha)^2 - v_0^2}.
\end{equation}

\item
Smaller values of $\alpha$ lead to a different morphology (\textit{composite bulge}) in which the mutant profile joins the flat portion at a finite slope $\sigma$. 
Close to the sector boundary, we can propose a traveling front solution of the form 
\begin{equation}\label{eq:front1}
h(x,t|f=1) = v_0 t + \phi(x-u t), \quad{\rm with}\quad \phi(0)=0.
\end{equation}
The function $\phi$ must  satisfy
\begin{equation}\label{eq:front2}
- u \phi' = \alpha + \frac{v_0}{2}(\phi')^2+\nu\phi''\,.
\end{equation}
In the vicinity of the sector boundary, this identifies the constant slope 
\begin{equation}\label{slope-u}
\sigma =\phi'\approx \frac{\sqrt{u^2-2\alpha v_0}-u}{v_0}.
\end{equation}
The above equation also identifies $\alpha_c=u^2/(2v_0)$ as the boundary to the circular arc region.
\item
Note that the slope in Eq.(\ref{slope-u}) becomes negative for $\alpha<0$. This leads to a V-shaped dent morphology depicted in Fig.~\ref{fig:morphologies}. Since $s_0>0$, the mutant still invades the faster growing native species, which may appear somewhat counterintuitive. However, such dented fronts were observed experimentally by Lee \emph{et al.}~\cite{Lee2021}, who explained them via geometric constructions. These morphologies had already been anticipated in earlier work coupling front shape and competition~\cite{HorowitzKardar2019}.
\end{itemize}

\begin{figure}[t]
\centering
\includegraphics[width=0.75\textwidth]{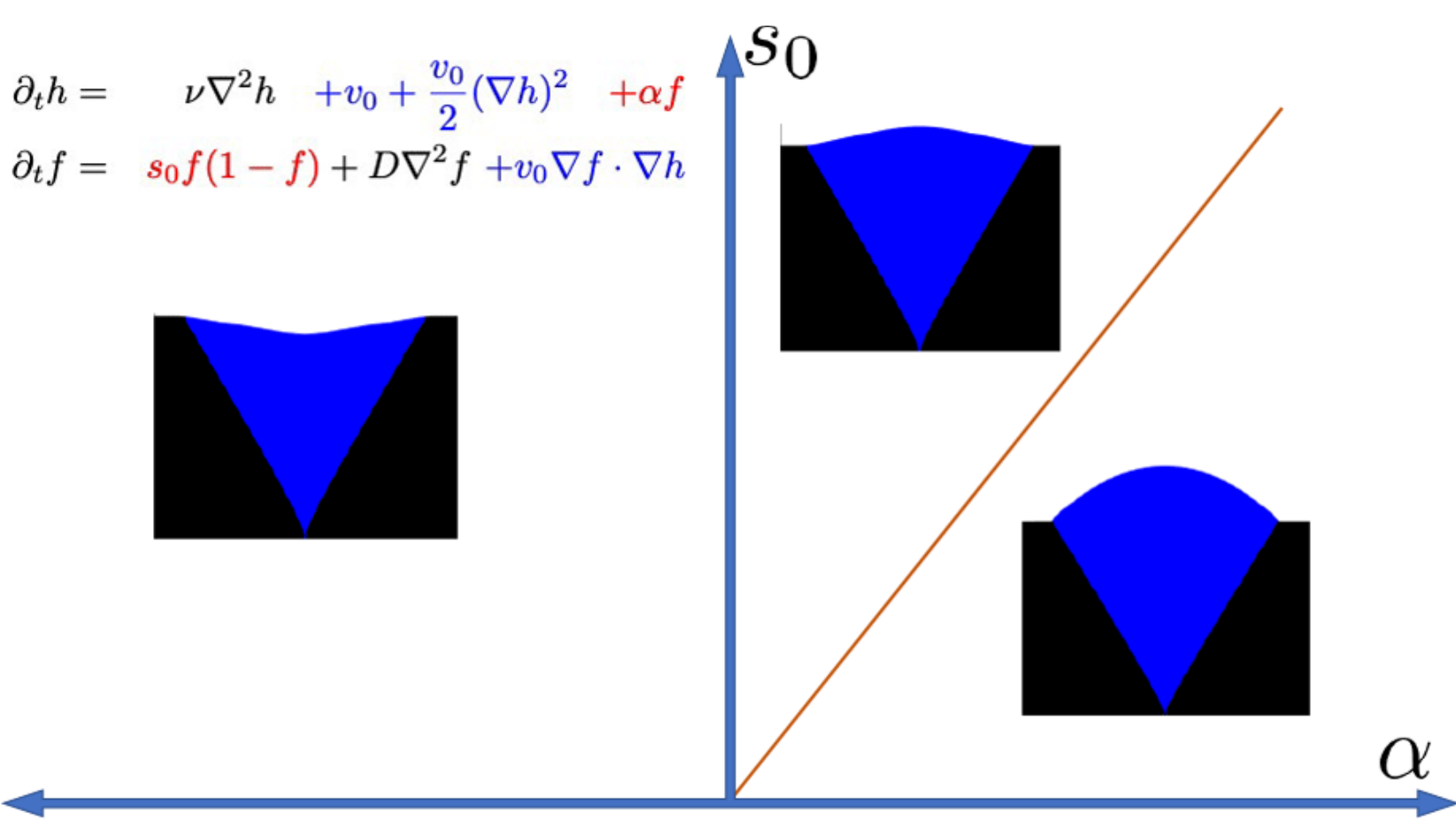}
\caption{Phase diagram of invasion morphologies for $s(f)=s_0f$. The (lateral) invasion velocity is set either by circular front geometry for $\alpha>\alpha_c=2Ds_0/v_0$ or by the Fisher velocity $\alpha<\alpha_c$.}
\label{fig:phase_diagram}
\end{figure}

Equation~(\ref{slope-u}) relates the slope $\sigma$ to an unspecified invasion velocity $u$. In performing numerical simulations of the coupled equation, we found that for $s(f)=s_0 f$ the invasion speed always equals $u=2\sqrt{s_0 D}$, which is the value corresponding to the uncoupled Fisher equation. This sets the transition between circular-arc and composite bulge morphologies at $\alpha_c=2Ds_0/v_0$, leading to the phase diagram depicted in Fig.~\ref{fig:phase_diagram}

\begin{figure}[t]
\centering
\includegraphics[width=0.75\textwidth]{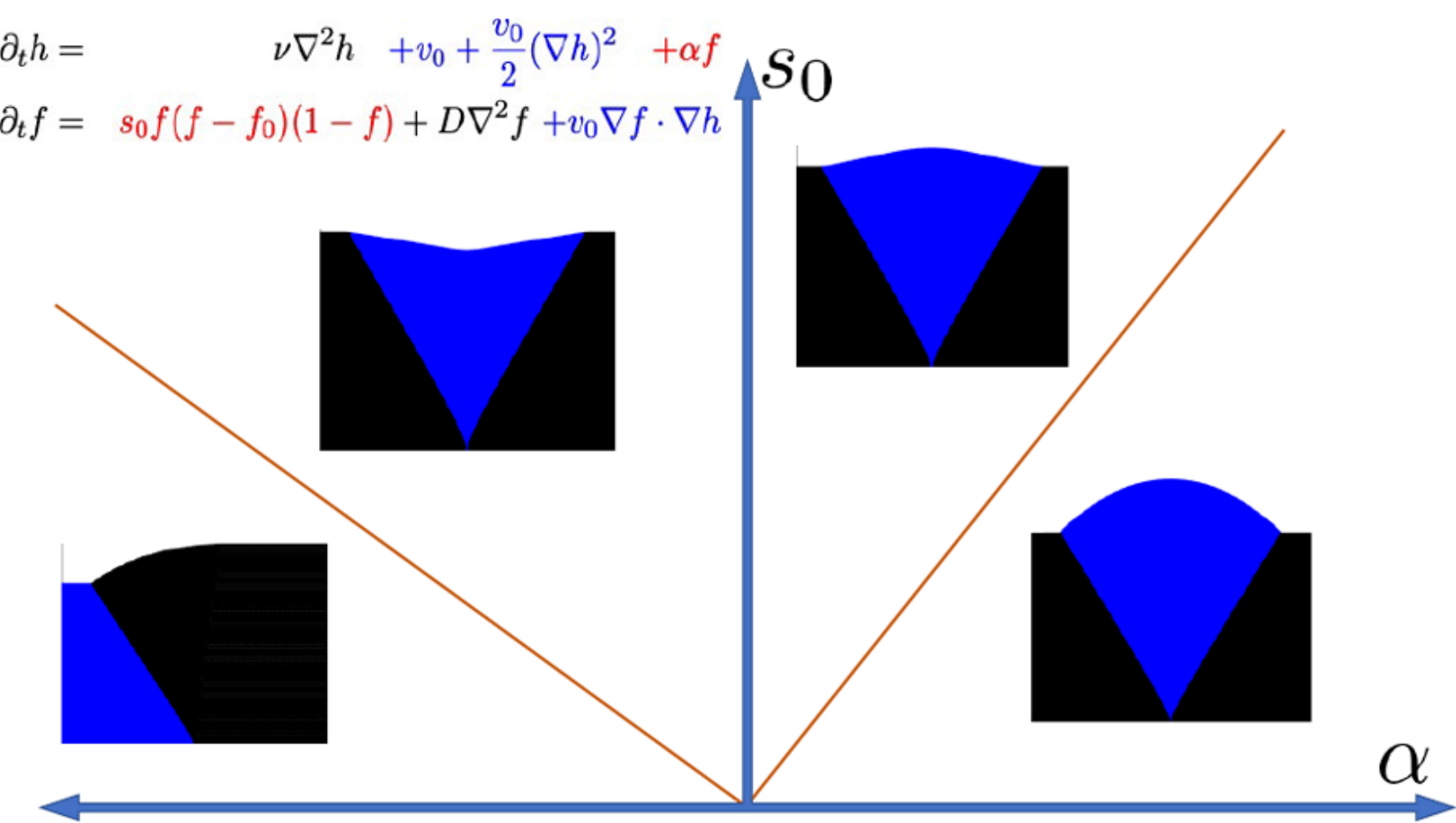}
\caption{Phase diagram of invasion morphologies with $s(f)=s_0f(f-f_0)$. For sufficiently negative $\alpha$ the mutant cannot invade the native species. }
\label{fig:phase_diagram2}
\end{figure}
 The lack of dependence of the Fisher velocity on the coupling to the profile is somewhat surprising. It turns out that this is a consequence of dealing with a \textit{pulled invasion front} for $s(f)=s_0f$. Richer behavior arises for \textit{pushed fronts}~\cite{vanSaarloos2003}, such as those generated by  $s(f)=s_0(f-f_0)$. This form appears in describing the  \textit{Allee effect} in which a minimal fraction $f_0$ is needed before the mutant acquires competitive advantage. This case, explored in~\cite{SwartzEtAl2023}, yields invasion velocities that vary continuously with $\alpha$ and can even become negative. This results in a new morphology, and the new phase diagram depicted in Fig.~\ref{fig:phase_diagram2}.

\section{Exact results from the Cole--Hopf transformation}

An important property of the KPZ equation is that it can be linearized by the Cole--Hopf transformation,
\begin{equation}
\partial_t h
=
\nu\nabla^2 h
+
\frac{\lambda}{2}(\nabla h)^2
+
\eta
\qquad
\Longleftrightarrow
\qquad
W=\exp\!\Big(\frac{\lambda h}{2\nu}\Big),
\label{eq:colehopf}
\end{equation}
which removes the nonlinearity of the dynamics at the expense of multiplicative $\eta$
\begin{equation}
\partial_t W=\nu\nabla^2 W+\eta W.
\label{eq:Wdiff}
\end{equation}
This transformation provides the precise link between the growing height $h(x,t)$ of the KPZ equation and the weight $W(x,t)$ of all directed paths in the random potential $\eta(x,t)$ ending at $(x,t)$. In the deterministic limit $\eta=0$, the mapping allows one to reconstruct exactly (using the diffusion kernel) the time evolution of the height profile for any initial condition $h(x,0)$.

We now ask whether a corresponding transformation exists for the coupled KPZ/Fisher-type equations introduced in the context of growth and competition. To address this question, we begin with a spatially extended version of the classic linear (Eigen) model of quasi-species evolution:
Consider a set of species indexed by $\alpha$, with local population densities $N_\alpha(x,t)$ that grow at rates $g_\alpha$, mutate at rates $\mu_{\alpha\beta}$, and all diffuse in space with the same diffusion coefficient $\nu$. In the deterministic limit, their evolution is governed by
\begin{equation}
\partial_t N_\alpha
=
g_\alpha N_\alpha
+
\sum_\beta\left(
\mu_{\alpha\beta}N_\beta
-
\mu_{\beta\alpha}N_\alpha
\right)
+
\nu\nabla^2 N_\alpha
\equiv
\hat L_{\alpha\beta}N_\beta.
\label{eq:eigen}
\end{equation}
Because the dynamics is linear, the formal solution is
\begin{equation}
N_\alpha(x,t)
=
\big[\exp(\hat L t)\big]_{\alpha\beta}
N_\beta(x,0).
\label{eq:eigenformal}
\end{equation}
However, such solutions generally grow or decay exponentially in time and must be interpreted with care as representing a population.

Summing Eq.~\eqref{eq:eigen} over all species yields the evolution of the total population density as
\begin{equation}
\partial_t N
=
\bar g\,N
+
\nu\nabla^2 N,
\quad{\rm where}\quad
\bar g(x,t)=\sum_\alpha g_\alpha f_\alpha(x,t),
\label{eq:totalN}
\end{equation}
in terms of  the local \textit{species fractions}
\begin{equation}
f_\alpha(x,t)=\frac{N_\alpha(x,t)}{N(x,t)},
\qquad
0\le f_\alpha\le 1,
\qquad
\sum_\alpha f_\alpha=1.
\end{equation}

The fractions $f_\alpha(x,t)$ evolve according to
\begin{equation}
\partial_t f_\alpha
=
(g_\alpha-\bar g)f_\alpha
+
\sum_\beta\left(
\mu_{\alpha\beta}f_\beta
-
\mu_{\beta\alpha}f_\alpha
\right)
+
\nu\left(
\nabla^2 f_\alpha
+2\nabla f_\alpha\cdot\nabla\ln N
\right).
\label{eq:falpha}
\end{equation}
The last term describes advection of the composition along gradients of $\ln N$, a structure closely reminiscent of the advection of sector boundaries along surface slopes encountered earlier in range expansions, Eq.~(\ref{eq:coupled_f}).

This observation motivates introducing a height-like field via
\begin{equation}
h(x,t)
=
\frac{2\nu}{\lambda}\,\ln N(x,t).
\label{eq:hlogN}
\end{equation}
Using this transformation, the height and the species fractions co-evolve according to
\begin{align}
\partial_t h
&=
\frac{2\nu}{\lambda}\,\bar g[f]
+
\nu\nabla^2 h
+
\frac{\lambda}{2}(\nabla h)^2,
\label{eq:h_f_general}
\\[4pt]
\partial_t f_\alpha
&=
(g_\alpha-\bar g)f_\alpha
+
\text{mutations}
+
\nu\nabla^2 f_\alpha
+
\lambda\nabla f_\alpha\cdot\nabla h\quad.
\label{eq:f_f_general}
\end{align}
These equations describe how the proposed height profile evolves according to the local composition of species, while gradients of the height advect  these local species fractions.

To make contact with the earlier discussion of two competing species in range expansion, we now restrict to two species with $f_1=f$ (mutant) and $f_2=1-f$ (native). Neglecting mutations and choosing growth rates $g_1=s_0$ and $g_2=0$, we obtain
\begin{align}
\partial_t h
&=
\frac{2\nu s_0}{\lambda}\,f
+
\nu\nabla^2 h
+
\frac{\lambda}{2}(\nabla h)^2,
\label{eq:h_two}
\\[4pt]
\partial_t f
&=
s_0 f(1-f)
+
\nu\nabla^2 f
+
\lambda\nabla f\cdot\nabla h\qquad.
\label{eq:f_two}
\end{align}
Equations~(\ref{eq:h_two})--(\ref{eq:f_two}) constitute a specific realization of the coupled KPZ/Fisher-type equations introduced earlier, with $D=\nu$ and a constrained coupling
\begin{equation}
\alpha=\frac{2\nu s_0}{\lambda}.
\label{eq:constraint}
\end{equation}

\subsection*{Exact consequences of the Cole--Hopf mapping}

We will now use this mapping to extract several exact and semi-exact results, and to explore some consequences of coupling growth, competition, and front geometry.

\paragraph{1. Transition between circular and sloped morphologies.}
For the pulled front described by Eqs.~(\ref{eq:h_two})--(\ref{eq:f_two}), two characteristic velocities appear. The Fisher velocity associated with the composition dynamics is
\begin{equation}
u_F = 2\sqrt{\nu s_0},
\end{equation}
while the circular-arc invasion velocity following from the height equation is
\begin{equation}
u_c = \sqrt{2\alpha v_0}.
\end{equation}
The transition between circular-arc morphologies and sloped fronts occurs when these two velocities coincide,
\begin{equation}
2\sqrt{\nu s_0} = \sqrt{2\alpha v_0}
\qquad \Longrightarrow \qquad
\alpha_c = \frac{2\nu s_0}{v_0}.
\label{eq:transitionline}
\end{equation}
Remarkably, this is precisely the constraint \eqref{eq:constraint} imposed by the Cole--Hopf mapping. Thus, the transformation maps exactly onto the transition line separating circular and sloped morphologies of the pulled front in Fig.~\ref{fig:phase_diagram}.

\paragraph{2. Invasion speed and logarithmic corrections.}
Consider initial conditions corresponding to an initial uniform native population $N_2(x,0)=1$,
with mutant species localized at the origin as $N_1(x,0)=M\delta(x)$.
The deterministic solution to Eq.~(\ref{eq:eigen}) then yields
\begin{equation}
N_1(x,t)=\frac{M\exp\!\left(s_0 t-\frac{x^2}{4\nu t}\right)}{\sqrt{4\pi\nu t}},
\qquad
N_2(x,t)=1.
\label{eq:Nsol}
\end{equation}
Defining the midpoint of the invading front by the condition $N_1(x_{1/2},t)=N_2(x_{1/2},t)=1$, we obtain
\begin{equation}
x_{1/2}(t)
\approx
2\sqrt{\nu s_0}\,t
-
\frac{1}{2}\sqrt{\frac{\nu}{s_0}}\,
\ln\left(\frac{4\pi\nu t}{M^2}\right).
\label{eq:bramsonlike}
\end{equation}
Thus, while the asymptotic front velocity is $2\sqrt{\nu s_0}$, the approach to this speed is logarithmically slow. This behavior is reminiscent of the well-known Bramson shift of Fisher fronts.
For standard pulled fronts, Bramson  predicts (for $\nu=s_0$) an asymptotic $-\frac{3}{2}\ln t$ correction. As emphasized by Derrida, at the transition between pulled and pushed fronts the coefficient crosses over from $-3/2$ to $-1/2$~\cite{vanSaarloos2003,Derrida2023}. Our result exhibits precisely such a $1/2$ prefactor, consistent with the fact that the exact solution is poised here at a transition  between circular and sloped invasion morphologies.
Moreover, this result generalizes naturally to radially expanding wavefronts in $d$ spatial dimensions, where the logarithmic correction acquires a prefactor of $d/2$.

\paragraph{3. Coarsening from structured initial conditions.}
The Cole--Hopf transformation enables analysis of more complex initial conditions, such as non-uniform seeding of species along an initially rough front $h(x,0)$—for example, an undulated knife-edge placed on the growth medium.
Starting from
\begin{equation}
N(x,0)
=\exp\!\left(\frac{\lambda}{2\nu}h(x,0)\right)
=
\sum_\alpha N_\alpha(x,0),
\end{equation}
the deterministic solution of the diffusion equation (absent mutations) gives
\begin{equation}
N(x,t)
=
\sum_\alpha
\int \frac{d^d x'}{(4\pi\nu t)^{d/2}}\,
\exp\!\left[
g_\alpha t
-
\frac{(x-x')^2}{4\nu t}
+
\frac{\lambda}{2\nu}H_\alpha(x')
\right],
\label{eq:generalN}
\end{equation}
where
\begin{equation}
H_\alpha(x)=
\begin{cases}
h(x,0), & \text{if species $\alpha$ is seeded at } x,\\
-\infty, & \text{otherwise}.
\end{cases}
\end{equation}
At long times, the integral is dominated by saddle points, leading to the asymptotic height profile
\begin{equation}
h(x,t)
=
\frac{1}{\lambda}
\max_{\alpha,x'}
\left[
\lambda\left(H_\alpha(x')-\frac{(x-x')^2}{2t}\right)
+
2\nu g_\alpha t
\right].
\label{eq:saddle}
\end{equation}

\begin{figure}[t]
\centering
\includegraphics[width=0.75\textwidth]{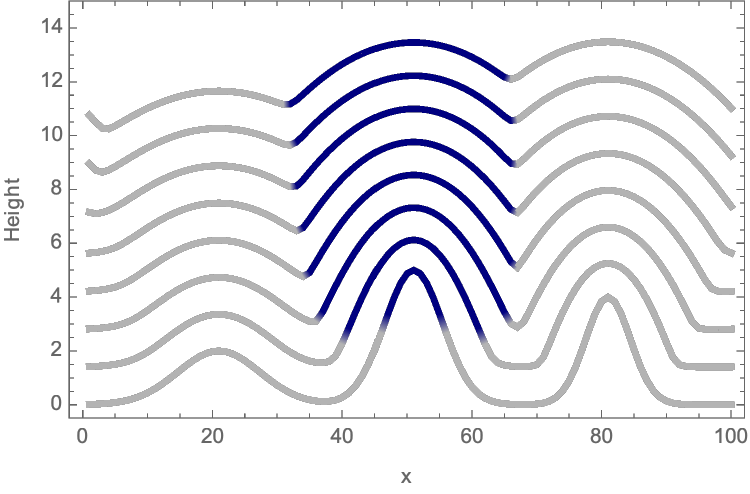}
\caption{Coarsening of an initially structured front. Each emerging paraboloid is dominated by a single species originating from a local peak in the initial profile.}
\label{fig:coarsening}
\end{figure}

As illustrated in Fig.~\ref{fig:coarsening}, an initially non-flat profile evolves into a collection of coarsening paraboloids, each composed of a single species descended from one seeded at an initial maximum. In particular, as shown in the example taken from Ref.~\cite{HorowitzKardar2019}, a less-fit (darker) species (which would go extinct on a flat front) can persist by exploiting a favorable initial geometric position. In this way, initial spatial heterogeneity creates long-lived geographic niches that stabilize otherwise disadvantaged species.

\paragraph{4. Growth or extinction of mutant on a sloped front.}

Next consider the fate of a localized mutant seeded on an initially sloped front. The mutant species is again localized to the origin as $N_1(x,0) = M\,\delta(x)$, but to introduce a profile with asymptotic slopes of $\pm s$, we set $N_2(x,0)=
\cosh[\lambda sx/(2\nu)]$ (corresponding to $h_0(x\to\pm\infty)\simeq s|x$).

As before, in the deterministic limit the time evolution of the linear equations is straightforward, with
\begin{equation}
N_1(x,t)
=
\frac{M}{\sqrt{4\pi \nu t}}\,
\exp\!\left(
s_0 t
-
\frac{x^2}{4\nu t}
\right),
\label{eq:N1_wedge}
\end{equation}
and
\begin{equation}
N_2(x,t)
=
\exp\!\Big(\frac{\lambda^2 s^2}{4\nu} t\Big)\,
\cosh\!\Big(\frac{\lambda s}{2\nu} x\Big).
\label{eq:N2_wedge}
\end{equation}

To determine whether the mutant invades the sloped region, we examine the locus where the two species are equally abundant, $N_1(x,t)=N_2(x,t)$, and ask whether this “front” propagates with a nonzero asymptotic velocity. Focusing on the $x>0$ side and using the large-$x$ approximation
\(
\cosh(a x) \simeq \tfrac{1}{2} e^{a x},
\)
we set $x=u t$ and equate (\ref{eq:N1_wedge}) and (\ref{eq:N2_wedge}) at large $t$:
\begin{equation}
\frac{M}{\sqrt{4\pi \nu t}}\,
\exp\!\left(
s_0 t
-
\frac{u^2 t}{4\nu}
\right)
\simeq
\frac{1}{2}\,
\exp\!\left(
\frac{\lambda^2 s^2}{4\nu} t
+
\frac{\lambda s}{2\nu} u t
\right).
\end{equation}
Taking logarithms and keeping only the leading terms in $t$ gives
\begin{equation}
s_0
-
\frac{u^2}{4\nu}
\simeq
\frac{\lambda^2 s^2}{4\nu}
+
\frac{\lambda s}{2\nu}u~.
\end{equation}
Multiplying by $4\nu$ yields a quadratic equation for the asymptotic front speed $u$:
\begin{equation}
u^2 + 2\lambda s\,u + \lambda^2 s^2 - 4\nu s_0 = 0~.
\label{eq:quad_u}
\end{equation}
Focusing on the positive $x$ region, the  relevant root is
\begin{equation}
u
=
-\lambda s
+
2\sqrt{\nu s_0}\,.
\label{eq:u_solution}
\end{equation}

Equation~\eqref{eq:u_solution} exhibits two distinct regimes, controlled by the comparison of the Fisher growth rate $s_0$ to the effective geometric scale $\lambda^2 s^2/(4\nu)$:

\smallskip
\noindent
\emph{Class A: mutant invasion ($u>0$).}  
If
\begin{equation}
s_0 > \frac{\lambda^2 s^2}{4\nu},
\end{equation}
then $u>0$ and the mutant continues to invade the sloped region at long times with a finite lateral velocity given by \eqref{eq:u_solution}. In the height perspective, a bulge dominated by the mutant develops near the origin and advances into the wedge, while the asymptotic slopes $\pm s$ are recovered only far from the invasion front.

\smallskip
\noindent
\emph{Class B: mutant extinction ($u\leq 0$).}  
If
\begin{equation}
s_0 \leq \frac{\lambda^2 s^2}{4\nu}\,,
\end{equation}
then $u\leq 0$ and the equal-abundance boundary drifts back toward the origin. The region occupied by the mutant shrinks and it becomes extinct.

\section{Fitness variations across an evolving population}

% This section is intended as an outline / scaffold
% for a fuller treatment of fitness distributions
% in evolving populations, to be completed in collaboration.

%\subsection*{Motivation: what does the fitness distribution look like?}

Let us pose the following question: if we could measure the fitness of all individuals in a large population at a given time, what would the resulting distribution look like? Would it be approximately Gaussian, or a more complicated shape that carries  signatures of the past history of expansions and competitions of the population?
%This question becomes particularly relevant if lineages can be tracked over many generations, suggesting that fitness distributions may reveal underlying universality and connections to stochastic growth processes.

The distribution of fitness across evolving populations has been studied extensively in well-mixed settings. 
For example, Ref.~\cite{tsimring_rna_1996} modeled RNA virus populations as distributions evolving through a one-dimensional fitness space, introducing low-dimensional mean field theories as viable alternatives to the high-dimensional sequence space. %presented by Wright in 1931 
 Subsequent work introduced a traveling-wave picture: in large asexual populations undergoing continuous adaptation, the fitness distribution propagates as a quasi-deterministic solitary wave toward higher fitness, with speed controlled by rare fluctuations at the stochastic edge of the distribution~\cite{desai_beneficial_2007, hallatschek_noisy_2011}. The resulting steady-state fitness distribution is approximately Gaussian in the bulk, reflecting central-limit-theorem averaging, though the tails are shaped by stochastic effects at the leading edge~\cite{good_distribution_2012}. 

These results establish a baseline for well-mixed populations: fitness fluctuations are relatively tame, and Gaussian statistics emerge in the bulk of the population from aggregation of weakly correlated increments. The situation may change qualitatively when spatial structure and range expansion are introduced, motivating the question of whether other universal distributions, such as the Tracy--Widom (TW) distribution~\cite{tracywidom1994airy}, 
may emerge in spatially structured populations.

%\subsection*{A baseline: the (non-competitive) mother machine}

\begin{figure}[t]
  \centering
  \includegraphics[width=1.0\textwidth]{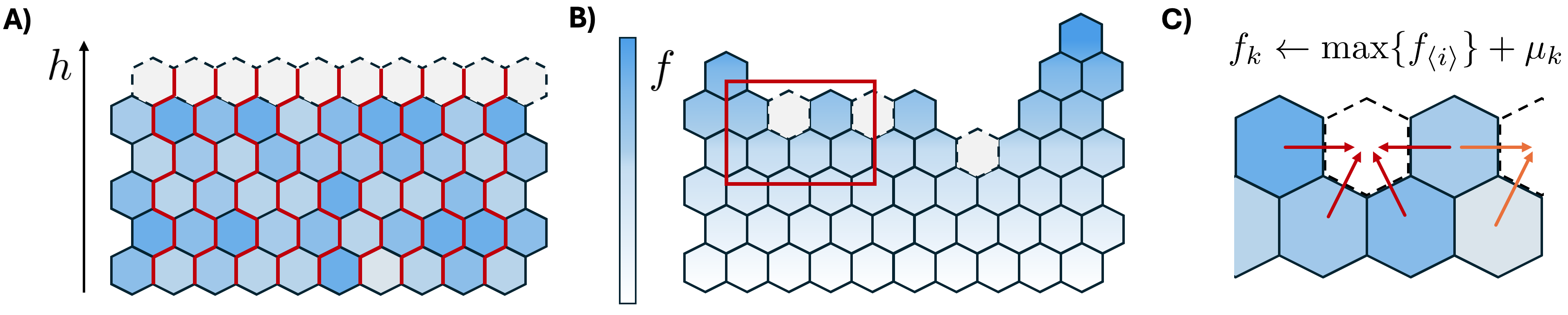}
\caption{(A) The mother machine geometry with synchronous reproduction leading to a flat expansion front. Red lines separate independent growth channels. (B) Competitive mother machine geometry with a rough front. Color indicates fitness. (C) Neighborhood competition rule: empty sites are colonized by the fittest individual. Arrows indicate potential parents competing for an expansion site.} 
\label{fig:mother}
\end{figure}

As a starting point, we take inspiration from the ``mother machine'' setup introduced in Ref.~\cite{wang_robust_2010} where $L$ growth channels (represented as corrugated columns in  Fig.~\ref{fig:mother}A) are arranged in parallel, each one oriented perpendicular to a shared front of reproducing cells. 
Now, consider the simple, and likely unrealistic, case of synchronous reproduction in each column, with  the height profile of the front remaining flat (for cells of identical size). 
If in each generation the descendant undergoes a small mutation, 
$f_i^{t+1}=f_i^t+\mu_i^t$, central-limit arguments predict an approximately Gaussian distribution of the accumulated fitness, with width that grows in time as $\sqrt{{\rm Var}f^t}\sim t^{1/2}$. 
%Additionally, in the limit of synchronous reproduction, the height profile of the front remains flat (for cells of identical size), and there is no correlation between the spatial location of individuals and their fitness.
%This simple setting serves as a baseline: In the absence of competition, and with synchronous growth,  
%that makes contact with the known results from the well-mixed literature: in the absence of strong spatial correlations or front roughening, fitness fluctuations are likely to be relatively straight forward, with Gaussian statistics emerging from aggregation of many weakly correlated increments.

%\subsection*{Range expansion and the connection to KPZ/Tracy--Widom}
The previously introduced stepping--stone models now enable us to consider how competition and morphology in range-expanding populations could lead to qualitative changes. As discussed in earlier sections,  individuals at the expansion front both shape the  front, and carry the lineages that dominate future generations. 
%The stochastic dynamics of the front height $h(x,t)$ is generically described by the KPZ universality class, whose fluctuations in one dimension are now known to be governed by Tracy--Widom distributions in many settings~\cite{corwin2012kpz}.
It is then reasonable to inquire in the
%A reasonable expectation is that, in range expansions, the 
fitness, $f$ of individuals at the front is somehow correlated to their position, e.g. to the height $h$ in a suitably defined growth profile. Through such correspondence, the statistics of fitness variations across the frontier can be coupled to the statistics of height fluctuations, which are now known to be governed by Tracy--Widom (TW) distributions in many settings~\cite{corwin2012kpz}. %suggesting that Tracy--Widom distributions may arise naturally in the distribution of fitness across the population.

Let $ \delta h(x,t) = h(x,t) - v_0 t $ indicate fluctuations at the front of a range expansion with average front speed $ v_0 $. A potential coupling arises if being ahead by $\delta h$ corresponds to an effective time advantage of $ \delta t = \delta h / v_0$. Thus, for individuals with identical growth rate $g$, the corresponding reproductive advantage due to spatial considerations is $\delta N = \exp(g \delta t)$. If an effective fitness can be associated with the log-abundance advantage as
\begin{equation}
  \label{eq:eff-fitness}
  \delta f_{\text{eff}} \equiv \log(\delta N/N) = g\delta t \propto \delta h\,,
\end{equation}
fitness fluctuations across the frontier reflect height fluctuations.
The latter random variables (on short time scales) have the form
\begin{equation}
  \delta h \sim \left(\Gamma t\right)^{1/3} \chi_{\beta}\,.
\end{equation}
where $\chi_{\beta}$ is Tracy-Widom distributed~\cite{corwin2012kpz}. The specific form of the fluctuations are known to depend on the initial condition of the height profile. For flat initial conditions, $h(x,0) = h_0$, height fluctuations obey the Gaussian Orthogonal Ensemble $(\beta = 1)$ TW distribution. On the other hand, a droplet initial condition $h(x,0) = -|x|/\delta$ with $\delta \ll 1$, yields fluctuations in the Gaussian Unitary Ensemble TW $(\beta = 2)$ distribution \cite{prahofer_universal_2000, takeuchi_growing_2011}.

\subsection*{The competitive mother machine model}

Motivated by the above arguments, we introduce a simple model where multiple individuals compete to colonize empty space. We consider a height field $h_i^t$ of an expanding population in 1+1 dimensions and evolve it according to the well-studied surface growth model known as the restricted solid-on-solid (RSOS) model~\cite{RSOS}. At each time step, a random site $k$ is chosen uniformly and is increased in height by one unit if the RSOS condition is satisfied,
\begin{equation}
    |h_k^t- h_{k\pm1}^t| \leq 1.
\end{equation}
This condition restricts the slope between adjacent sites to be at most one, disallowing any overhangs so that $h_i^t$ is a well-defined function. A possible configuration of the height profile is depicted in Fig.~\ref{fig:mother}B. 

The competitive mother machine is defined as a variant of the standard RSOS model by also considering a fitness field $f_i^t$ in addition to the height field. After updating the height field at a site $k$, the fitness at the same site is updated according to a nearest-neighbor competition rule,
\begin{equation}
    f_k^{t+1} = \max\{f_{\langle i\rangle}^{t}\}+\mu_k^t\,.
\end{equation}

The nearest neighbors are the occupied sites that make direct contact with the site to be occupied and are depicted in Fig.\ref{fig:mother}C. The descendant thus inherits the fitness of the winning parent as well as an added mutation $\mu_i^t \sim \mathcal{N}(0,\sigma^2)$. While the height function evolves as in the standard RSOS model, the evolution of the fitness field is constrained by the height, since only sites that are allowed to grow can have their fitness value updated. Thus, the height and fitness evolve as coupled stochastic fields whose properties we study via computer simulation.
%\subsection*{Scaling of Fluctuations and Fitness Distribution}

\begin{figure}[t]
    \centering
    \includegraphics[width=1\linewidth]{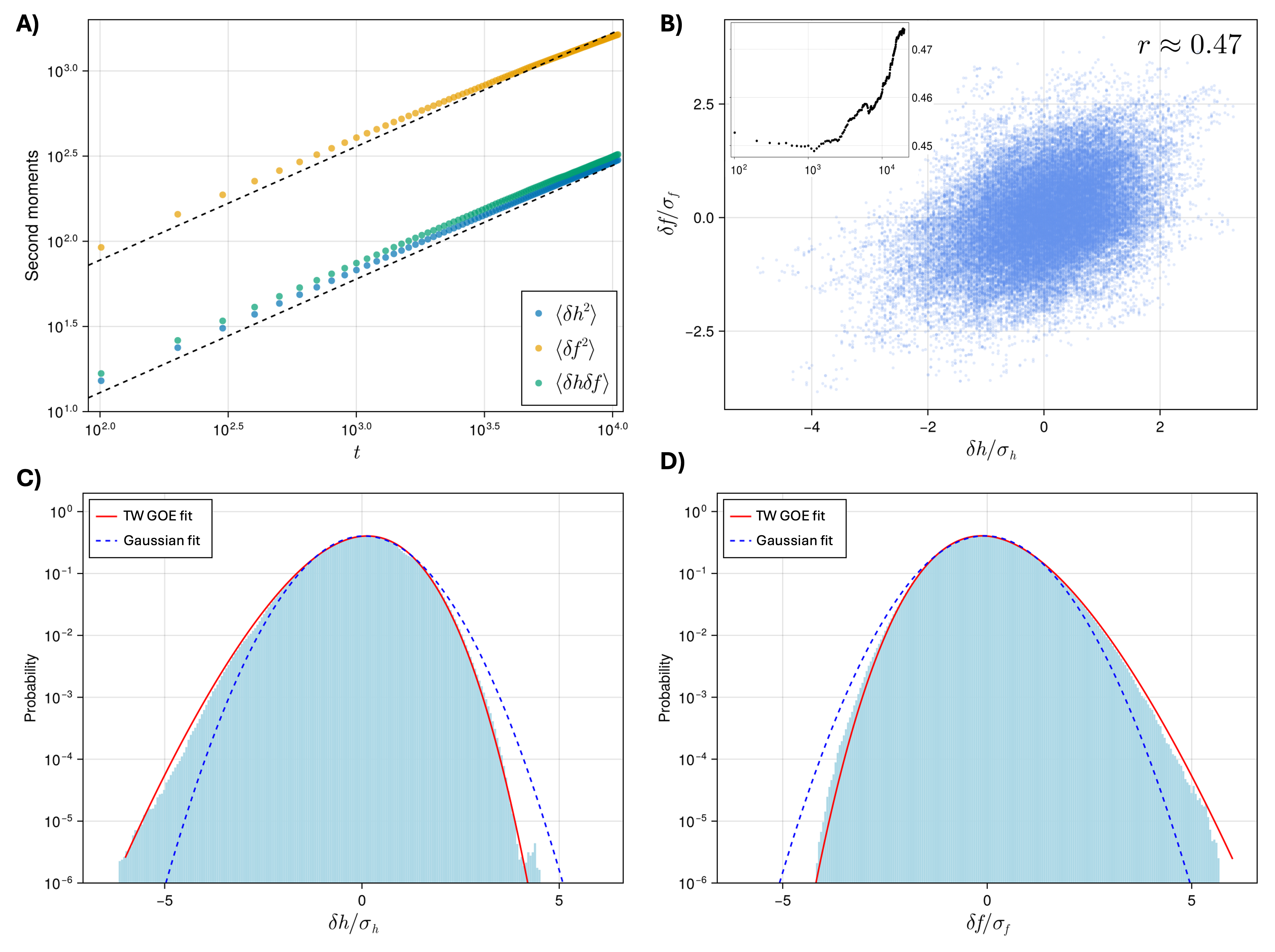}
    \caption{A) Numerical support for the KPZ scaling of second moments of $h,f$; dashed lines have slope $2/3$. B) The height and fitness fluctuations are positively correlated. Inset: Pearson correlation $r(t)$ over simulation runtime. C) Distribution of the height fluctuations of the leading front; and D) fitness fluctuations at leading front. (Both fitted to TW distributions after centering and scaling.)}
    \label{fig:TW}
\end{figure}

Figure~\ref{fig:TW}  presents numerical results for simulations of the competitive mother machine model. We measure over $R = 10^4$ realizations of size $L = 10^4$ for a runtime of $t = L$, with flat initial conditions and periodic boundary conditions.
We measure the variance of height fluctuations $\text{Var}(h) = \langle \delta h^2 \rangle$, the variance of fitness fluctuations $\text{Var}(f) = \langle \delta f^2 \rangle$, and their covariance $\text{Cov}(h,f) = \langle \delta h \, \delta f \rangle$ , where averages are taken over the front and over independent realizations. As depicted in  Fig.~\ref{fig:TW}A, all three quantities exhibit the characteristic KPZ scaling, growing as $t^{2/3}$. A distinctive feature of this system is the nontrivial correlation between height and fitness, which is quantified by a correlation coefficient that hovers around $r \sim +0.47$ (Fig.~\ref{fig:TW}B). This positive correlation reflects the expectation that advancing portions of the front have won more local competitions and thus accumulated higher fitness.

Furthermore, unlike the well-mixed case with a distribution that is approximately as Gaussian in the bulk, in the competitive mother machine model both height and fitness are better described by a Tracy-Widom GOE distribution (Fig.~\ref{fig:TW}C and D). We note that the height distribution displays negative skew due to the competition rule and growth channel geometry which constrains the growth of the interface to local valleys rather than peaks. In other words, upon coarse-graining, one  obtains a negative non-linear $(\lambda < 0)$ coefficient in the KPZ equation for the RSOS height profile. The fitness distribution, by contrast, exhibits positive skew as expected from the selection rule which systematically favors high-fitness lineages.

We note that the role of front roughening can be isolated by considering the synchronous limit, in which all sites advance together and the front remains flat, $h_i^t \propto t$ for all $i$ (Fig.~\ref{fig:mother}A). In this limit, height fluctuations and height-fitness correlations vanish while the fitness distribution remains Tracy-Widom (not shown). %due to the max-mutation rule. 

%\subsection*{Summary}

The competitive mother machine model thus demonstrates that spatially structured, range-expanding populations may generically exhibit non-Gaussian fitness statistics. In contrast to well-mixed models where central-limit averaging yields Gaussian distributions in the bulk, the combination of local competition and front roughening produces Tracy-Widom distributed fluctuations in both height and fitness, providing a qualitative departure from well-mixed theory. These predictions are experimentally accessible. Microbial range expansions on agar plates already exhibit KPZ-consistent front roughening \cite{Hallatschek2007}; measuring fitness distributions across the front could reveal height-fitness correlations. Overall, we conclude that studying fitness variations in spatially structured, range-expanding populations could offer a natural bridge between classical questions in evolutionary biology and modern developments in KPZ/Tracy--Widom universality. 

\subsection{Acknowledgments}
MK acknowledges long-time support by the NSF, most recently through grant DMR-2218849. SE acknowledges support by the MIT Dean of Science fellowship.

\clearpage

\bibliographystyle{iopart-num}
\bibliography{kpz_references}
\end{document}